# Phonon driven non-linear electrical behaviour in molecular devices

Antonino La Magna and Ioannis Deretzis

*CNR - IMM, Stradale Primosole 50, 95121 Catania, Italy*

**Abstract**

Electronic transport in a model molecular device coupled with local phonon modes is theoretically analyzed. The method allows for obtaining an accurate approximation of the system's quantum state irrespectively from the electron and phonon energy scales. Non-linear electrical features emerge from the calculated current-voltage characteristics. The quantum corrections with respect to the adiabatic limit characterize the transport scenario, and the polaronic reduction of the effective device-lead coupling plays a fundamental role in the unusual electrical features.

Significant experimental and theoretical advances [1-7] in the electron transport study at a molecular level, achieved in the last decade, are driven by the possibility to complement the actual Si-based nanoelectronics with molecular-based ones. One of the intriguing aspects of prototype molecular devices is their "intrinsic" highly non-linear character, showing Negative Differential Resistance (NDR) or hysteresis. Indeed, non-linearity of molecular devices is not an added functionality but it seems related to fundamental features due to the electron and/or electron-vibration interactions. A careful evaluation of the effects of the electron-phonon scattering is at the basis of the prediction of linear electron transport features when semi-classical [8], semi-quantum [9] and full-quantum [10] schemes are applied. These schemes stem from the assumption that the electron-phonon coupling is weak and a pertubative approach is reliable. However, in a molecular device the vibration can couple strongly with electrons, and only a non-pertubative study of the models [6] can support the understanding of phonon driven unusual electron dynamics.

Making good contacts between leads and the active device portion is maybe the hottest issue in molecular electronics [4,7]. Indeed, it is unlike that the experimental realisation of a molecular device results in an ideal contacting: the coupling between the metal and the molecular states can be either weak or strong in dependence on the nature of the metal, on the interface chemical bonding etc. In the case of relative weak coupling with the contact, the electron and phonon energy scales become comparable, making also questionable the use of the Born-Oppenheimer (BO) adiabatic approximation in the studied electron-phonon interactions.

The aim of this work is to study the phonon driven non-linear electrical behaviour of an ideal molecular device using an accurate approximation of the system's quantum state irrespectively of the electron and phonon energy scales. Our results indicate an origin of the non-linear character of molecular devices that has not been considered in previous interpretations.

As schematisation of the molecular device Hamiltonian, we consider a single spinless electron level coupled to a single vibrational mode and to the leads. The Hamiltonian [11] reads

$$H = \varepsilon_0 c_0^+ c_0 + \hbar\omega a^+ a + \chi(a^+ + a) c_0^+ c_0 + \sum_{k \in \{L,R\}} \varepsilon_k c_k^+ c_k + \sum_{k \in \{L,R\}} (V_k c_k^+ c_0 + h.c.) \quad (1)$$

where $\varepsilon_0$ is the molecule electron energy level, $\omega$ the frequency of the phonon mode, $\chi$ the electron-phonon coupling energy, $\varepsilon_k$ the energy of the electron states in the two leads $L$ and $R$, and $V_k$ the device-leads couplings. Note that $\varepsilon_0$ can be tuned in the case of a three terminal device configuration. An optimal variational state for the electron-phonon Hamiltonian has to include the competitive character of limit solutions in the adiabatic ($\hbar\omega \ll V$) and anti-adiabatic ($\hbar\omega \gg V$) limits, considering the effects of both static (adiabatic) and dynamic (anti-adiabatic) distortions; moreover we should also take into account the role of the anomalous (non-gaussian) fluctuation of the phonon state characterising the intermediate ($\hbar\omega \approx V$) regime [12,13]. Defining $\lambda = \chi^2/\hbar\omega$, static distortion is introduced by means of the translation transformation

$$U_1 = \exp(-S_1) = \exp\left[-\sqrt{\frac{\lambda}{\hbar\omega}}(a^+ - a)\tilde{x}_0\right],$$

where $\tilde{x}_0$ is the dimensionless distortion, which on the basis of the variational principle (see below) is equal to the electron density in the level. Similarly the dynamic distortion is considered by means of a Lang-Firsov [5,14] type transformation

$$U_2 = \exp(-S_2) = \exp\left[-\sqrt{\frac{\lambda}{\hbar\omega}}\vartheta(a^+ - a)(c_0^+ c_0 - \tilde{x}_0)\right]$$

where $\vartheta$ measures the weight of the Small Polaron (SP) anti-adiabatic character of the solution. We introduce the anomalous fluctuations averaging $H_2 = U_2^{-1} U_1^{-1} H U_2 U_1$ with a squeezed phonon state

$$|\Phi_{ph}\rangle = \exp\left[-\alpha(aa - a^+ a^+)\right]|0_{ph}\rangle$$

where $\alpha$ is a measure of the displacement from the standard gaussian fluctuations for the phonon quantum state. The resulting pure electron effective Hamiltonian is

$$H_{eff} = \tilde{\varepsilon}_0 c_0^+ c_0 + \sum_{k \in \{L,R\}} \varepsilon_k c_k^+ c_k + \sum_{k \in \{L,R\}} (\tilde{V}_k c_k^+ c_0 + h.c.) + \lambda \tilde{x}_0^2 (1-\vartheta)^2 + \frac{\hbar\omega}{4}(\tau^2 + \tau^{-2}) \quad (2)$$

where

$$\tilde{\varepsilon}_0 = \varepsilon_0 - \lambda + \lambda(1-\vartheta)^2(1-2\tilde{x}_0); \quad \tilde{V}_k = \exp\left[-\frac{\lambda}{2\hbar\omega}\vartheta^2\tau^2\right]V_k; \quad \tau = \exp(-2\alpha)$$

Note that the BO and SP solutions are obtained for $\vartheta = 0; \tau = 1$ and $\vartheta = 1; \tau = 1$ respectively. The extension of the phonon-mediated correlation does not have to be considered for our model, while it could be important for a multi-level or/and a multi-phonon model [15]. At zero temperature $T$ and bias $V_{bias}$ the Ground State (GS) energy is [16]

$$E_{GS} = \frac{1}{\pi}\int_{-\infty}^{\mu}\tan^{-1}\left(\frac{\tilde{\Delta}(E)}{E-\tilde{\varepsilon}_0-\Lambda(E)}\right)dE + \lambda\tilde{x}_0^2(1-\vartheta)^2 + \frac{\hbar\omega}{4}(\tau^2+\tau^{-2}) \quad (3)$$

where $\mu$ is the chemical potential and

$$\Delta_{L,R}(E) = 2\pi\sum_{k\in\{L\},\{R\}}|V_k|^2\delta_{Dirac}(E-\varepsilon_k); \quad \tilde{\Delta}(E) = \Delta(E)\exp\left[-\frac{\lambda}{2\hbar\omega}\vartheta^2\tau^2\right]; \quad \Lambda(E) = \frac{1}{\pi}P\int_{-\infty}^{\infty}\frac{\Delta(E)}{E-E'}dE'$$

where $P$ indicates the principal value integral and $\Delta(E) = 0.5\times(\Delta_L(E)+\Delta_R(E))$. If we define

$$n = \frac{1}{\pi}\int_{-\infty}^{\mu}\frac{\tilde{\Delta}(E)}{[E-\tilde{\varepsilon}_0-\Lambda(E)]^2+\tilde{\Delta}(E)^2}dE \quad (4)$$

$$S = \frac{1}{\pi}\int_{-\infty}^{\mu}\frac{\tilde{\Delta}(E)[E-\tilde{\varepsilon}_0-\Lambda(E)]}{[E-\tilde{\varepsilon}_0-\Lambda(E)]^2+\tilde{\Delta}(E)^2}dE \quad (5)$$

where n is the electron density state and $S$ is an energy shift, the extreme conditions for $E_{GS}$ are

$$\frac{\delta E_{GS}}{\delta\tilde{x}_0} = 0 \quad \Rightarrow \quad \tilde{x}_0 = n = \frac{1}{\pi}\int_{-\infty}^{\mu}\frac{\tilde{\Delta}(E)}{[E-\tilde{\varepsilon}_0-\Lambda(E)]^2+\tilde{\Delta}(E)^2}dE \quad (6)$$

$$\frac{\delta E_{GS}}{\delta\vartheta} = 0 \quad \Rightarrow \quad \vartheta = \frac{(1-n)n}{(1-n)n-\frac{S}{2\hbar\omega}\tau^2} \quad (7)$$

$$\frac{\delta E_{GS}}{\delta\tau^2} = 0 \quad \Rightarrow \quad \tau^2 = \sqrt{\frac{1}{1-\frac{2\lambda S}{(\hbar\omega)^2}\vartheta^2}} \quad (8)$$

Expressions $\Delta(E)$ and $\Lambda(E)$ depend on the particular bands of the leads (i.e on $\varepsilon_k$ and $V_k$); however, we can derive the general behaviour of the system in the Wide Band Limit Approximation (WBLA) where $\Delta(E) = \Delta$, $\Lambda(E) = 0$ and introducing a lower negative cut-off $-W$ for the contact bands in the integral expression for $S$ [17]. In this limit we find analytic expressions for $n$ and $S$

$$n = \frac{1}{2} - \frac{1}{\pi}\tan^{-1}\left[\frac{\tilde{\varepsilon}_0 - \mu}{\tilde{\Delta}}\right]; \quad S = \frac{\tilde{\Delta}}{2\pi}\log\left[\frac{(\tilde{\varepsilon}_0 - \mu)^2 + \tilde{\Delta}^2}{(\tilde{\varepsilon}_0 + W)^2 + \tilde{\Delta}^2}\right] \quad (9)$$

The optimal *GS* estimate for intermediate values of the model parameters shows features quite different with respect to the limit (SP or BO) solutions. In fig. 1 we show the analysis of the *GS* solution for the following set of parameters: $W = 12, \Delta = 1, \lambda = 3, \varepsilon_0 = 3$ (here and in the following we use $\hbar\omega$ as units for the energy). The global minimum for $E_{GS}$ has been obtained for $n = \tilde{x}_0 = 0.5; \ \vartheta = 0.5202; \ \tau = 0.6933$. From the comparison between fig.1a and fig.1b we can derive that our best *GS* estimate does not show either the bi-stable character (as a function of *n*) of the BO solution [6] or the lack of dependence on *n* of the SPs. We note that our $E_{GS}$=-0.199 estimate is significantly better than the BO one (the global minimum of $E_{GS}$ is −0.032 in this limit).

In fig. 2 we show as a function of $\varepsilon_0$ the solution corresponding to stable minima of the variational equations for a different set of the triples $W, \Delta, \lambda$. For each case, the stable minima in the BO limit are also shown. For all the sets of parameters here considered, the BO solutions show a bi-stable behaviour in the $(\lambda \approx \varepsilon_0)$ region while the optimal *GS* solution shows bi-stability in a restricted region of the parameters. Indeed, considering Figs.2 a-c, we note that our GS estimate is bi-stable only when $W, \Delta \gg \hbar\omega$ i.e when approaching to the BO limit. The increase of $\lambda$ favours bi-stability, as it can be inferred from the comparison between fig. 2.c and 2.d. However, the range of $\varepsilon_0$ where the best *GS* solution manifests bi-stability, is, in general, strongly reduced with respect to the one found in the BO limit (see e.g. Figs 2a and 2d).

The generalisation of the equations at $T \neq 0$ and $V_{bias} \neq 0$ is

$$n = \frac{1}{2} - \frac{1}{\pi}\tan^{-1}\left[\frac{\tilde{\varepsilon}_0 - E_L}{\tilde{\Delta}}\right] + \frac{1}{2\pi}\int_{E_L}^{E_U}\frac{\tilde{\Delta}_L f(E, \mu_L) + \tilde{\Delta}_R f(E, \mu_R)}{[E - \tilde{\varepsilon}_0]^2 + \tilde{\Delta}^2}dE \quad (10)$$

$$S = \frac{\tilde{\Delta}}{2\pi} \log\left[\frac{(\tilde{\varepsilon}_0 - E_L)^2 + \tilde{\Delta}^2}{(\tilde{\varepsilon}_0 + W)^2 + \tilde{\Delta}^2}\right] + \frac{1}{2\pi} \int_{E_L}^{E_U} \frac{[E - \tilde{\varepsilon}_0](\tilde{\Delta}_L f(E, \mu_L) + \tilde{\Delta}_R f(E, \mu_R))}{[E - \tilde{\varepsilon}_0]^2 + \tilde{\Delta}^2} dE \qquad (11)$$

where $\mu_L = \mu + 0.5\, V_{bias}$ and $\mu_R = \mu - 0.5\, V_{bias}$ are the contact electrochemical potentials, $f(E,\mu) = \{1 + \exp[\beta(E - \mu)]\}^{-1}$ is the Fermi-Dirac distribution; the lower and upper cut-off $E_L$ and $E_U$ are chosen as $f(E,\mu) \cong 1$ for $E < E_L$ and $f(E,\mu) \cong 0$ for $E > E_U$. We can also evaluate the device current by means of the Landauer formula

$$I = \frac{2e}{h} \int_{E_L}^{E_U} T(E)(f(E,\mu_L) - f(E,\mu_R)) dE = \frac{2e}{h} \int_{E_L}^{E_U} \frac{[\tilde{\Delta}_L \tilde{\Delta}_R]}{[E - \tilde{\varepsilon}_0]^2 + \tilde{\Delta}^2}(f(E,\mu_L) - f(E,\mu_R)) dE. \qquad (12)$$

In the non-equilibrium case a functional analogous to the *GS* energy whose local minima determine the stability condition does not exist. However, following Ref. [6], we can individuate the outer root (as function of the level occupancy *n*) of the Eqs. (6-8,11,12) as the locally stable ones. Moreover, increasing the bias an additional stable middle root has to be included among the locally stable ones (as we will see, for high bias this root is the single solution of Eqs. (6-8)).

In fig. 3 the occupancy levels estimated for the locally stable solutions as functions of the applied bias are shown for $W = 20, \Delta_L = \Delta_R = 1, \lambda = 6.5, \varepsilon_0 = 6.4, T = 50K$. In this case both optimal (solid lines) and BO (dashed lines) solutions have a bistable character at equilibrium ($V_{bias}=0$), however the behaviour of *n* as a function of $V_{bias}$ is completely different in the two cases. This fact and the correspondent different behaviour of the corrected coupling function $\tilde{\Delta}$ and of the shifted energy level $\tilde{\varepsilon}_0$ cause a different estimate of the *I(V)* characteristics (fig. 4).

According to the optimal solution predictions, the system has a larger conductance near the $V_{bias}=0$. This fact is related to a weaker localisation (not compensated by the reduction of $\tilde{\Delta}$) predicted by the optimal solution with respect to the BO one. Hysteresis cycles could be figured out both in the optimal and in the BO *I(V)* estimates, however the optimal solution predicts cycles at a lower bias with smaller current jumps. The most striking evidence is the NDR predicted by the optimal solution for $V_{bias} > 2.1\ V$. NDR occurs also in the BO solution when the shifted energy level crosses the window between the chemical potentials of the leads [6]. However, the NDR in the optimal solution is related to a different mechanism showing a more stringent non-linear character of the device in the intermediate regime of the parameters. Indeed, (see inset in fig. 4) $\tilde{\varepsilon}_0$ is almost constant in the $V_{bias} > 2.1$ Volts region while $\tilde{\Delta}_L$ and $\tilde{\Delta}_R$ decrease, tending to the SP solution for large $V_{bias}$. *Therefore, in this case NDR is related to a polaronic effect which weakens the coupling between the leads and the molecular device.* This effect cannot be evidenced in the BO limit, used in Ref. [6], where $\tilde{\Delta} = \Delta$ does not effectively depend on the electron-phonon interaction.

Although our model captures noticeable physical features, the approximations here considered deserve a discussion. The scenario presented is not significantly affected by the WBLA. Indeed, if we consider, for example, a tight-binding contact band $\varepsilon_k = -W\cos(k)$ and constant device-leads coupling $\Delta_{TB} = \Delta$, the calculated *I(V)* characteristics show a similar qualitative behaviour (see the magenta line in fig. 4). The coupling between the active phonon mode and bath phonons could effectively reduce the strength of the electron-phonon correlation contrasting the polaron formation. In a first approximation we can consider the bath role re-normalising $\lambda \to \lambda \omega^2 /[\omega^2 + (\gamma/2)^2]$ where γ is a parameter related to the phonon-phonon interaction [6]. This re-scaling of λ should be explicitly considered when the coupling with the bath is not negligible. Moreover, non equilibrium phonon's dynamics could characterise the dissipation of the device through the bath. This phenomenon has been studied in the SP limit [18]; however the inclusion of the quantum correction to the active phonon state could modify the results obtained in Ref. [18].

Our approach can be easily generalised to take into account the spin $\sigma = \pm 1/2$ and electron correlation (adding the on site Hubbard term $Uc_{0,\sigma}^+ c_{0,\sigma} c_{0,1-\sigma}^+ c_{0,1-\sigma}$). This generalisation enriches the scenario described since the phonon mediated electron correlation $U_{eff} = U - \lambda(2\vartheta - \vartheta^2)$ can be attractive or repulsive [15]; and different regimes can be established as a function of $U, \lambda, \Delta$. At the Hartree-Fock level, variational equations are similar to Eqs. (6-8), considering two levels $\widetilde{\varepsilon}_{0\sigma} = \varepsilon_0 - \lambda(2\vartheta - \vartheta^2) + [U - \lambda(2\vartheta - \vartheta^2)]n_{1-\sigma} - 2\lambda(1-\vartheta)^2 n_\sigma$ and the replacement of $n(n-1)$ with $n(n-0.5)$ in Eq. 7 ($n = n_{1/2} + n_{-1/2}$ and $0 < n < 2$). Our calculations show that a similar polaronic mechanism for the NDR is recovered also in this case, as shown in fig. 4 (green line) where we also plot the *I(V)* curve obtained adding the electron correlation (*U=7*).

In conclusion, this work, based on an accurate polaron transport theory, shows that in the $\hbar\omega \approx V$ regime of the electron-phonon coupling the molecular device's quantum state has a proper response to the applied potential. As a consequence we have evidenced a mechanism for the non-linear electronic behaviour of the bridge related to the "potential dependent" renormalization of the effective coupling constant due to the polaronic effect. This fact can open new perspectives in the interpretation of the constantly growing experimental evidences of non-linearity in molecular devices, especially for systems with active redox centres [2,3,19-22]. Indeed, the origin of the non-linearity is not well understood and the various *ad hoc* mechanisms proposed result in a re-organisation of the device's levels; while the polaronic re-organisation of the couplings with the contacts, indicated by our approach, has never been considered. Of course, the application of the model to experimental systems requires the correct parameter calibration and the eventual extension to the multilevel case.

**Figure captions**

**Fig.1** Ground state energy estimate as a function of the variation parameters $n, \theta$ (upper panel) for a fixed value of $\tau^2 = 0.6933$. The model parameters are $W = 12, \Delta = 1, \lambda = 3, \varepsilon_0 = 3$. b) Ground state energy estimate in the adiabatic limit as a function of $n$ (lower panel) for the same set of model parameters.

**Fig.2** Optimal variational parameter estimates as functions of the unpertubated $\varepsilon_0$ molecular energy level for different sets of $W, \Delta, \lambda$. Black lines indicate $n$, blue lines $\vartheta$ and red lines $\tau^2$. The estimate of the level filling $n$ in the adiabatic limit is also shown as a dashed black line.

**Fig. 3** Level filling for the different stable solutions as a function of the applied bias $V_{bias}$ estimated by means of our method (solid line) and using the adiabatic approximation (dashes). The parameters set are $W = 20, \Delta_L = \Delta_R = 1, \lambda = 6.5, \varepsilon_0 = 6.4$. Outer roots are plotted in black and blue, inner roots in red.

**Fig. 4** Current/Voltage characteristics of the different stable solutions estimated by means of our method (solid line) and using the adiabatic approximation (dashes). The parameter are $W = 20, \Delta_L = \Delta_R = 1, \lambda = 6.5, \varepsilon_0 = 6.4$. Outer roots are plotted in black and blue, inner roots in red. Outer root for V>0 obtained for the same set of parameter but avoiding the wide band limit and including electron correlation (U=7) are shown as magenta and green solid line respectively. In the inset we show the coupling constant reduction $\tilde{\Delta}_L/\Delta_L = \tilde{\Delta}_R/\Delta_R$ (solid line) and the shifted energy level $\tilde{\varepsilon}_0$ (dashes) as a function of the applied bias $V_{bias}$>0 for the inner optimal stable solution.

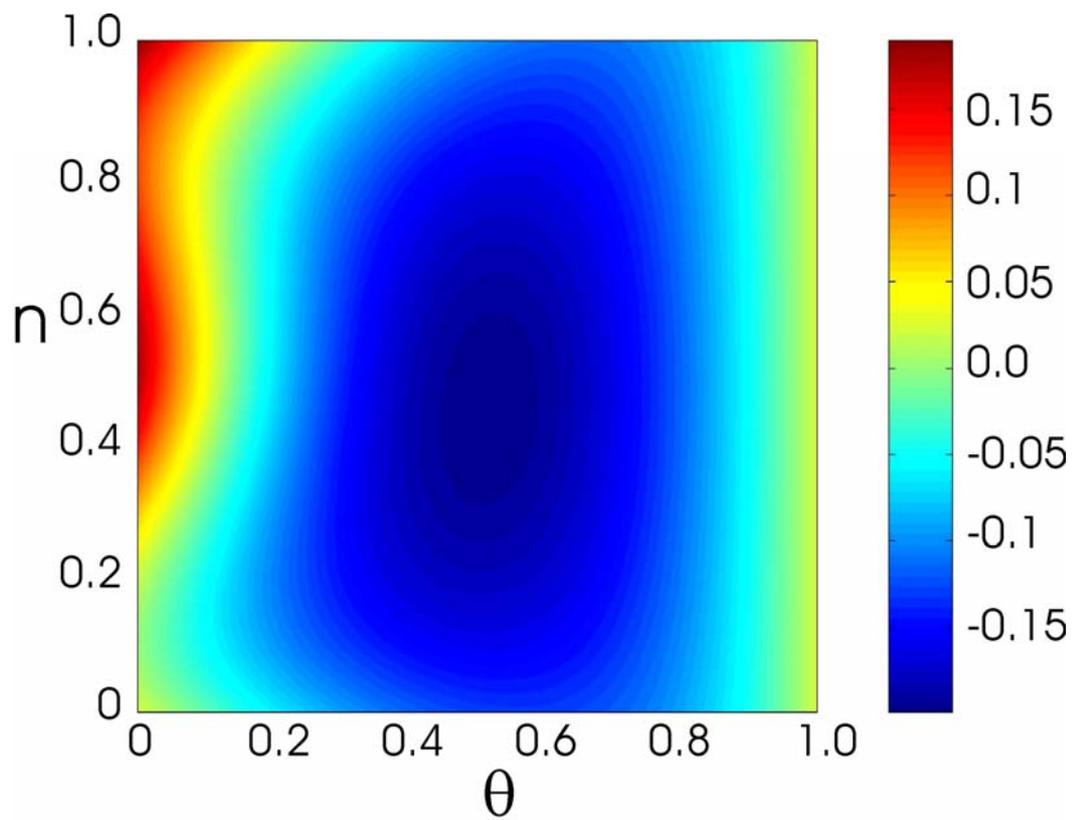
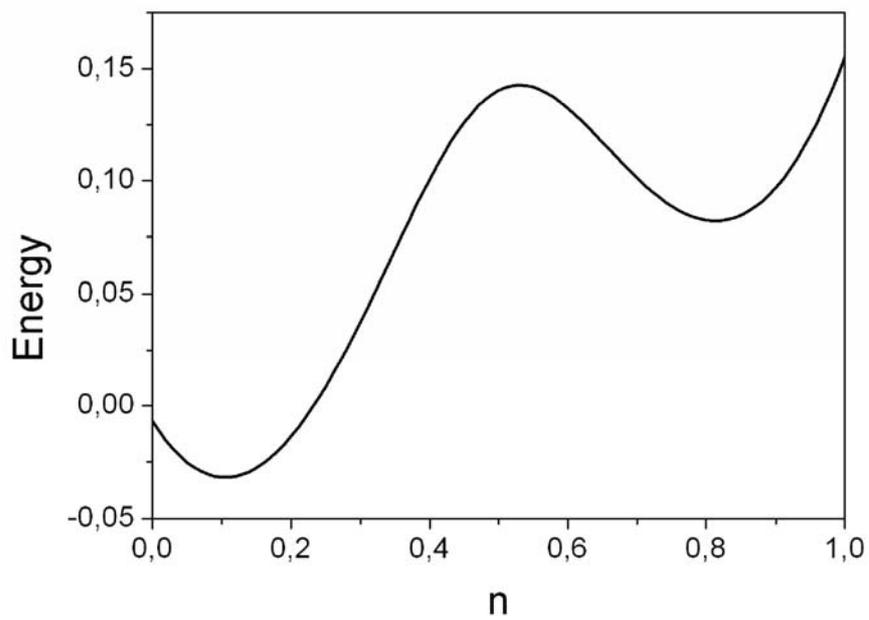

A. La Magna and I. Deretzis Fig. 1/4

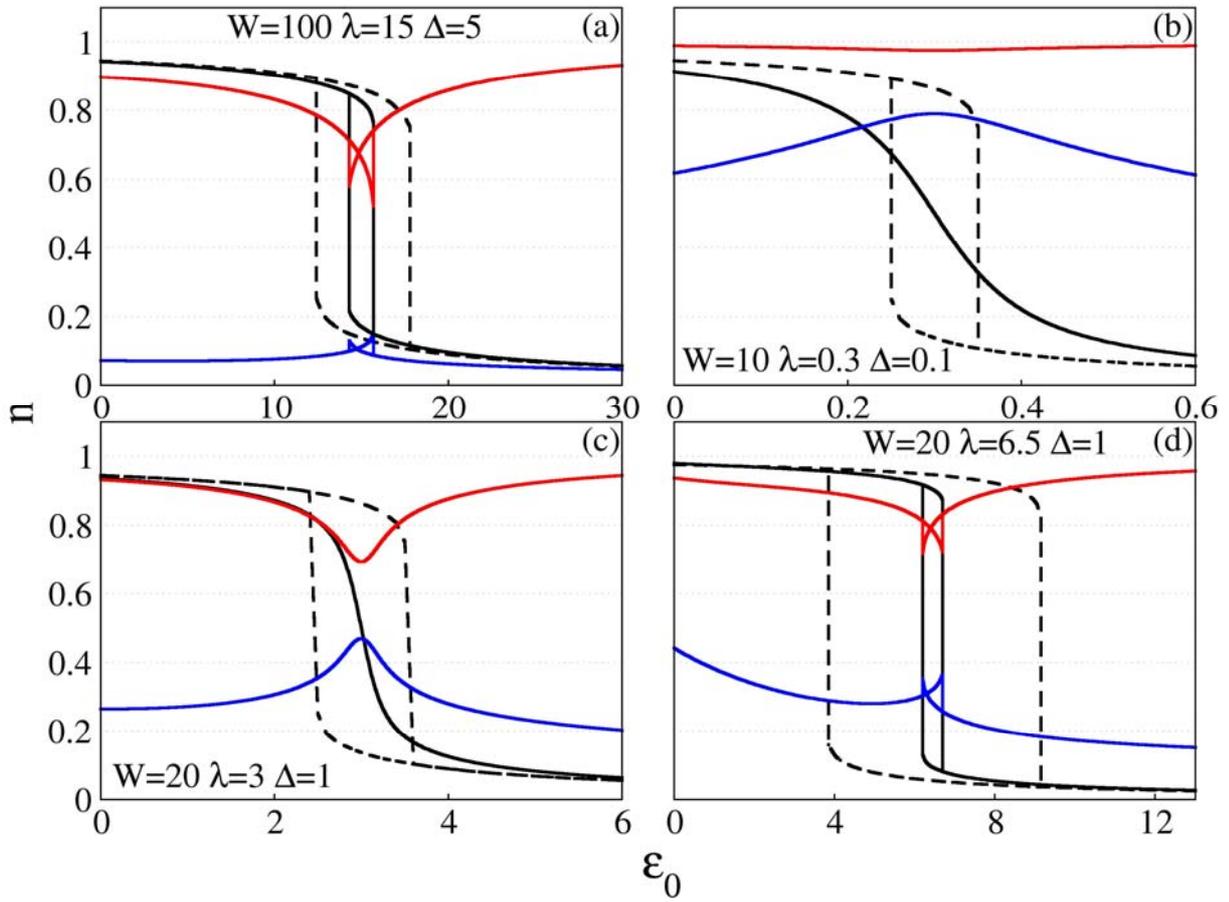

A. La Magna and I. Deretzis Fig. 2/4

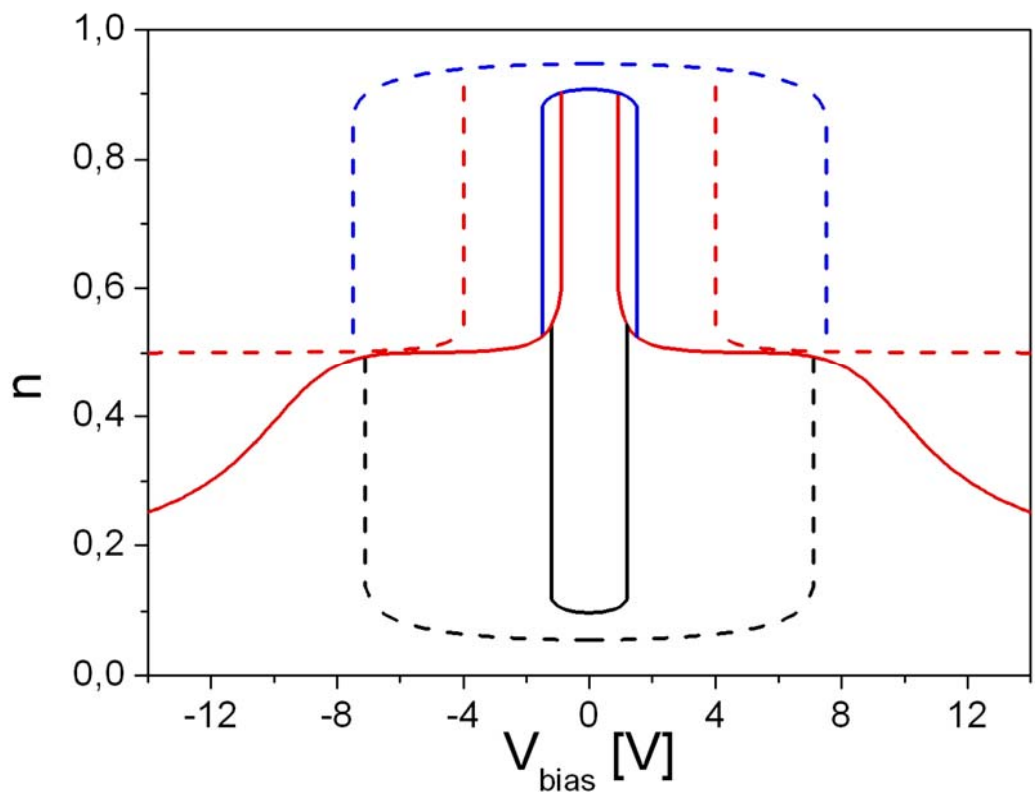

A. La Magna and I. Deretzis Fig. 3/4

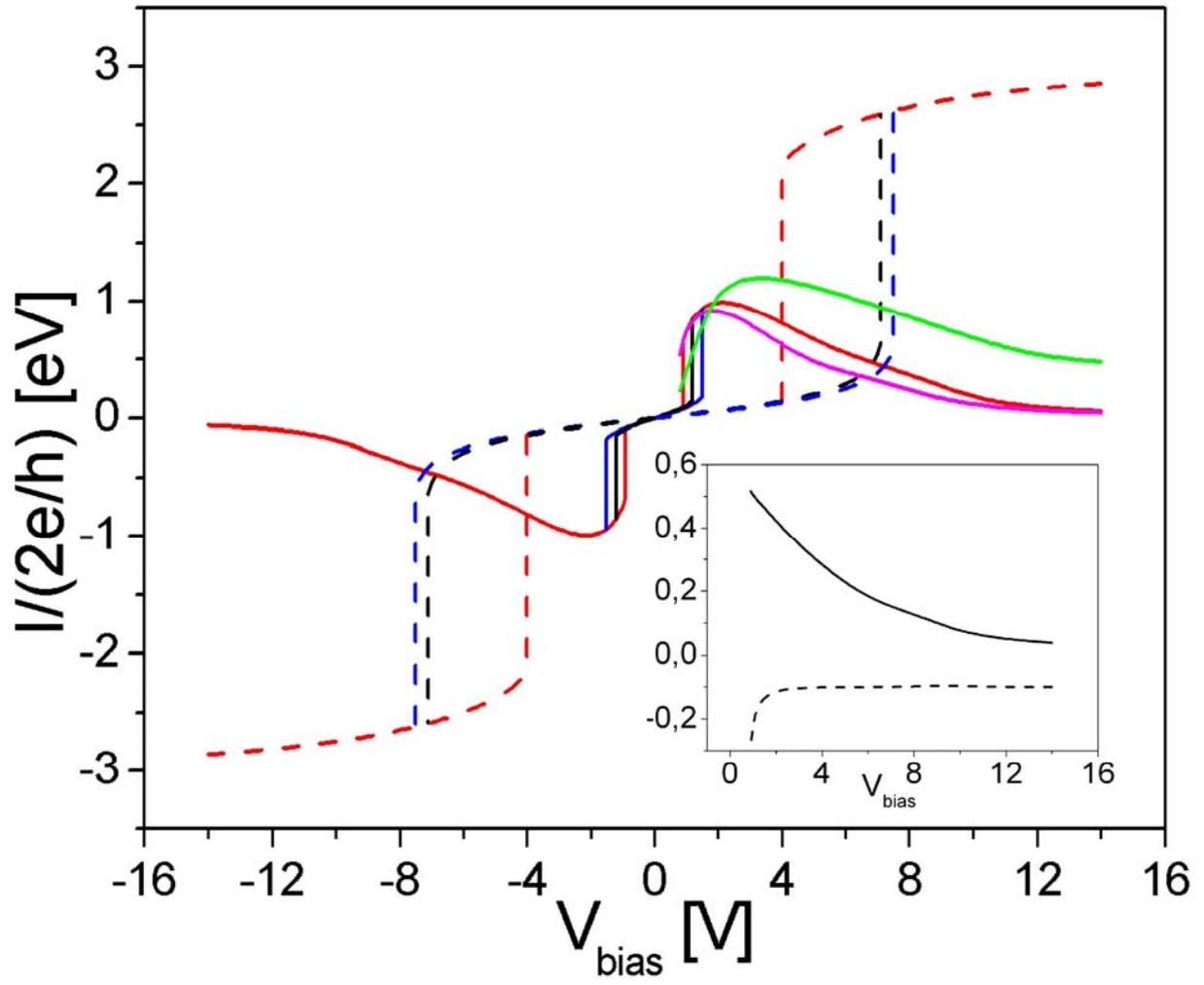

A. La Magna and I. Deretzis Fig. 4/4